\documentclass[aps,pre,floats,floatfix,twocolumn]{revtex4}

\usepackage{graphicx}
\usepackage{amssymb} 
\usepackage{color}
\usepackage{hyperref}

\graphicspath{{./}{./Figs/}}

\def\bc{\begin{center}}
\def\ec{\end{center}}

\newcommand{\beq}{\begin{eqnarray}}
\newcommand{\eeq}{\end{eqnarray}} 
\newcommand{\dn}{\downarrow}
\newcommand{\up}{\uparrow}

\newcommand{\ket}[1]{\left| #1 \right\rangle}

\newcommand{\Braket}[2]{ \left\langle #1 \middle| #2 \right\rangle}
\newcommand{\BraKet}[3]{ \left\langle #1 \middle| #2 \middle| #3 \right\rangle}

\newcommand{\rmrk}[1]{{#1}} 
\newcommand{\rmnt}[1]{}  

\newcommand{\trc}{\mbox{Trace}}
\newcommand{\hrefl}[1]{\href{#1}{[link]}}

\newcommand{\Eq}[1]  {{\textcolor{blue}{Eq.}}~(\ref{#1})} 
\newcommand{\Fig}[1] {{\textcolor{blue}{Fig.}}~\ref{#1}}

\begin{document}

\title{Hilbert-space localization in closed quantum systems}

\author{Doron Cohen$^{1}$, Vyacheslav I. Yukalov$^{2}$, Klaus Ziegler$^{3}$}

\affiliation
{
$^{1}$Department of Physics, Ben-Gurion University of the Negev, P.O.B. 653, Beer-Sheva 84105, Israel\\
$^2$Bogolubov Laboratory of Theoretical Physics, Joint Institute for Nuclear Research, Dubna 141980, Russia\\
$^3$Institut f\"{u}r Physik, Universit\"{a}t Augsburg, D-86135 Augsburg, Germany
}

\begin{abstract}
Quantum localization within an energy-shell of a closed quantum system
stands in contrast to the ergodic assumption of Boltzmann, and to the 
corresponding eigenstate thermalization hypothesis. The familiar case 
is the real-space {\em Anderson localization} and its many-body Fock-space version.  
We use the term {\em Hilbert-space localization} in order to emphasize 
the more general phase-space context. Specifically, we introduce a unifying 
picture that extends the semiclassical perspective of Heller, 
which relates the localization measure to the probability of return.   
We illustrate our approach by considering several systems of experimental interest, 
referring in particular to the Bosonic Josephson tunneling junction. 
We explore the dependence of the localization measure on 
the initial state, and on the strength of the many-body interactions 
using a novel recursive projection method. 
\end{abstract}

\pacs{42.50.Ar, 42.50.Pq, 42.50.Ct}

\maketitle

\section{Introduction}

For long time, the most important example of a complex \rmnt{added ``complex"} finite quantum system 
was the atomic nuclei \cite{Blaizot_86}. Nowadays, there exist several other types
of complex finite quantum systems that are experimentally accessible and widely 
studied, such as photon cavities, quantum dots, trapped atoms, metallic and 
magnetic nanoclusters, and graphene flakes 
\cite{cqed,Walther_06,Lipparini_08,Pethick_08,Yukalov09,Katsnelson_12,Birman_13}.
The existence of a variety of finite quantum systems opens a wide field for
studying fundamental quantum properties. In recent years, high interest has 
been directed to the investigation of such interconnected problems as 
equilibration and thermalization \cite{PSSV_11,Yukalov_11}, 
and entanglement \cite{Williams_98,Nielsen_00,Vedral_02,Keyl_02}. 

\rmrk{
Random scattering in a quantum or classical wave-like system can cause a severe interference 
effect that creates a complex dynamics. While we expect diffusion in the presence of very
weak random scattering, strong random scattering can lead to Anderson localization. In this
case a particle cannot escape from a finite region, defined by the localization length.
This picture was proposed in the seminal work of Anderson under the assumption that a single
quantum particle is scattered in a static random environment \cite{anderson58}. A natural
extension is a quantum gas, consisting of many particles. The description of such a system
is substantially more complex, requiring a many-body wave function rather than a single-particle
wave function in real space. The definition of localization is also different in the
many-body system because the relevant space is the many-body Hilbert space rather than the
real space of a single particle. On the other hand, a static random environment is not
necessary to produce interference because interparticle scattering plays a similar role:
An individual particle inside the quantum gas experiences scattering by other
particles. Since the dynamics of the gas is quite complex, the scattering of an individual 
particle by other particles can be considered as random. Then the main difference in comparison
to Anderson's picture is that the scattering environment is dynamic rather than static.  
Thus, we expect that the motion of the many-body system is constraint to subregions in
the Hilbert space by scattering events which cause strong interference of the many-body 
wave function. This will be called {\em Hilbert-space localization} subsequently, in contrast
to Anderson localization.  
Our approach should be distinguished from other many-body generalizations, where the 
interplay of disorder and interactions was addressed \cite{Basko_06,gora,Pal_10,Huse_13}.
} 

\rmrk{The first work that placed the Quantum-localization theme in the context 
of finite complex systems concerns ``the Standard Map", 
aka ``the Quantum Kicked Rotator model" \cite{QKRc}.
It has been realized \cite{QKRf} that the observed localization can be mapped 
to the one-dimensional Anderson model with quasi-random disorder. 
Thus, the underlying ``chaos'' induces an effective disorder, 
but the localization is not in space but in momentum, hence termed {\em dynamical localization}.
Subsequent studies have expanded this perspective. In particular we note  
the analysis of localization for coupled rotors \cite{TIPb}, 
which has been motivated by the interest in getting a better understanding for 
the coherent propagation of interacting particles in random potential \cite{TIPs,TIPb,TIPi}.}   

\rmrk{An important finding of Borgonovi and Shepelyansky \cite{TIPb} is the enhancement
of the localization length for two kicked rotators, as compared to the length of
a single kicked rotator. This result suggests that particle interactions can induce
an increased localization length or even
delocalization of otherwise localized single noninteracting particles, although in 
many other cases, particle interactions strengthen localization \cite{Kramer}. Thus,
interaction can produce both effects of either strengthening, weakening or even destroying localization.  
Another examples are cold atoms, where Anderson localization of noninteracting atoms
in random or quasiperiodic optical lattices can be destroyed by atomic
interactions (see review \cite{Modugno}).
}

\rmrk{Generally speaking, the role of interactions is not as simple, and they can 
produce both effects, either enhancing localization or destroying it. Their influence
depends on details of the considered physical system, including Bose or Fermi
statistics, the peculiarity of the energy spectrum, the specifics of the interaction
forces, whether the latter are short-ranged or long-ranged, repulsive or attractive,
and thermodynamic characteristics, such as temperature or pressure, also play their
role.}

\rmrk{We would like to point out that in a {\em small} complex system, 
disorder is in general not a relevant notion, and the localization effect
can depend in a non-monotonic way on the strength of the interactions.
A minimal example for that is the observed localization 
in the 4~site Bose-Hubbard model~\cite{ckt}. 
What matter are the characteristics of the phase-space structure.
The interplay between disorder and interaction becomes an issue 
if one considers larger clusters, still the same framework should 
handle all cases on {\em equal footing}.}

\rmrk{In the present paper we address the question of quantum localization 
from a more general point of view. Using the term Hilbert-space localization (HSL) 
we want to make clear that quantum localization does not have to show up 
in a particular dynamical variable. Our perspective is motivated by the work of Heller \cite{Heller_87} 
regarding the semiclassical picture of weak localization in phase-space, 
aka ``scarr theory". Here we extend this perspective and use it to discuss {\em strong localization},  
irrespective of whether it originates from disorder or from interactions between particles, 
and irrespective of whether it is in ``position" or in ``momentum".}
\\ 

The outline of the paper is as follows. In section~II, we distinguish between two 
notions of phase-space exploration. This allows for a generalized definition of quantum 
breaktime in sections~III and ~IV, which is the reason for having quantum localization. 
In Section~V, we discuss the calculation of the localization measure, while in Section~VI, 
we illustrate the procedure with regard to the Bosonic Josephson junction.  
\rmrk{For completeness, the traditional notion of spatial localization 
and its related entropy-based measures are briefly summarized 
in Appendix~A and in the Appendix B. The other appendices contain  
models-related material that has not been included in the main text.}  
\\

\section{Exploration of phase-space}

In ``Quantum localization and the rate of exploration of phase space" 
\cite{Heller_87} Heller has provided a semiclassical perspective for quantum 
localization of eigenstates. His framework was effective for the discussion 
of weak localization and scarring, while the strong localization effect, as 
well as the many-body localization theme, were left out of the semiclassical 
framework. 
We would like to refine the phase-space semiclassical framework in order to 
achieve a more comprehensive heuristic understanding of quantum localization. 
This proposed extension incorporates the dynamical breaktime concept 
(see \cite{dittrich,brk} and further references therein) that had been 
introduced in order to shed light on the strong Anderson localization effect 
in ${d=1,2,3}$ dimensions. Our starting point is a distinction between two 
different notions of participation numbers: 
\beq \nonumber
\mathcal{N}_{\text{cells}}(t) &=& \mbox{\#explored ``cells" at time~$t$} \; ,
\\ \nonumber
\mathcal{N}_{\text{states}}(t) &=& \mbox{\#participating states up to time~$t$} \; .
\eeq
We shall define these two functions below. Schematic illustration of them in the 
case of a diffusive-like system is provided in \Fig{fig:NvsT}.
For the purpose of the present section, it is useful to have in mind a simpler example: 
the free expansion of a wavefunction is a chaotic ballistic (non diffusive) billiard. 
The initial state is a Gaussian wavepacket. Two energy scales are involved:
\beq
\Delta_0 =& \mbox{mean level spacing} &\equiv \ {2\pi\hbar}/{t_H} \\
\Delta_E =& \mbox{energy shell width} &\equiv \ {2\pi\hbar}/{\tau_E}  \; ,
\eeq
where we have defined corresponding time scales $t_H$ (Heisenberg time) 
and $\tau_E$ (uncertainty time). The width of the energy shell is determined by the 
energy uncertainty of the preparation. The effective Hilbert space dimension is 
\beq
\mathcal{N}_E \ \ = \ \ \frac{\Delta_E}{\Delta_0} \; .
\eeq
Without loss of generality, and for presentation clarity, we assume that all the out-of-shell 
states have been truncated and, hence, we regard $\mathcal{N}_E$ as the actual Hilbert space 
dimension. It follows that any ``participation number" $\mathcal{N}$ is a-priori bounded by 
the value $\mathcal{N}_E$.

Given an arbitrary statistical state $\rho$, the number of participating 
(pure) states $\mathcal{N} \equiv \exp[\mathcal{S}]$ can be defined either via 
the Shanon entropy $\mathcal{S} = - \trc[\rho \ln \rho]$, or via 
$\mathcal{N} = 1/\trc[\rho^2]$, or more generally via the Renyi entropy 
which involves $\trc[\rho^{\gamma}]$, see App.\ref{app:S} for definitions.       
An initial Gaussian wavepacket preparation ${\rho = |\psi\rangle\langle\psi|}$, 
as well as the evolved state ${\rho = |\psi(t)\rangle\langle\psi(t)|}$, 
have zero entropy, meaning that the number of participating states in any 
instant of time is $\mathcal{N}(t)=1$.
We now turn to explain the definitions of $\mathcal{N}_{\text{cells}}(t)$ 
following Boltzmann, and of $\mathcal{N}_{\text{states}}(t)$ following Heller. 

The procedure of Boltzmann is to divide phase space into cells. In the quantum 
version, one can define a corresponding partitioning of Hilbert space using 
a complete set of projectors $\hat{P}_n=|n\rangle\langle n|$. 
Then, from the coarse-grained distribution ${p_n}$, 
using any of the above Renyi measures, one can define 
the number of explored ``cells" $\mathcal{N}_{\text{cells}}(t)$. 
We shall prefer below to use the $\trc[\rho^2]$ based definition, Namely, 
\beq
&& \mathcal{N}_{\text{cells}}(t) \ = \ \left[ \sum p_n^2  \right]^{-1},
\\
&& \ \ \ \text{where} \ \ \  p_n(t) = \left|\Braket{n}{\psi(t)}\right|^2  \; . 
\eeq

Following Heller, we define also $\mathcal{N}_{\text{states}}(t)$, which was 
described in \cite{Heller_87} as the ``number of phase-space cells accessed". 
We prefer the term ``number of participating states up to time~$t$". We shall 
see in a moment that the semantics is important. The definition 
of $\mathcal{N}_{\text{states}}(t)$ is based not on the coarse grained 
$\rho$ at time~$t$, but rather on the time-averaged $\rho$ during the time 
interval ${[0,t]}$. Namely, 
\beq
\mathcal{N}_{\text{states}}(t) 
&=& \trc\left[  \left(\frac{1}{t}\int_0^t \rho(t') dt'\right)^2 \right]^{-1} \\
&=& \left[ \frac{2}{t} \int_0^t \left(1-\frac{\tau}{t} \right) \mathcal{P}(\tau) d\tau  \right]^{-1} \; .
\eeq   
The last equality relates $\mathcal{N}_{\text{states}}(t)$ to the survival 
probability. The latter is defined as follows: 
\beq \label{e8}
&& \mathcal{P}(t) \ = \ \left|\Braket{\psi(0)}{\psi(t)}\right|^2  
\ = \ \sum_{\alpha,\beta} p_{\alpha}p_{\beta} \, e^{i(E_{\beta}-E_{\alpha})t}, \ \ \ \\ 
&& \ \ \ \text{where} \ \ \  p_{\alpha} = \left| \Braket{E_{\alpha}}{\psi} \right|^2 \; .
\eeq
The function $\mathcal{N}_{\text{states}}(t)$ provides the size of the basis 
that is required in order to simulate the dynamics up to time~$t$. 
It is a-priori bounded by the value $\mathcal{N}_E$. Its asymptotic value is 
\beq  \label{e10}
\mathcal{N}_{\infty}
\ = \ \mathcal{N}_{\text{states}}(\infty)
\ = \ \left[ \overline{\mathcal{P}(t)} \right]^{-1} 
\eeq
where the overline indicated time averaging.
For a non-degenerate spectrum it follows that 
\beq \label{e11}
\mathcal{N}_{\infty} 
\ = \ \left[\sum p_{\alpha}^2\right]^{-1}
\eeq
This is known as the {\em participation number} (PN) of eigenstates.
The function $\mathcal{N}_{\text{states}}(t)$ can be regarded
as a time-dependent generalization of the conventional PN notion.

\section{The localization measure}

The concept of spatial localization can be adopted 
to more complex quantum systems \cite{Heller_87,Logan_90} 
by considering a general basis of states $\{|n\rangle\}$ 
that define ``locations" within the energy shell. 
Then the evolution starts with an initial state $|\psi\rangle$ 
and evolves in time as ${|\psi(t)\rangle = \exp(-i\mathcal{H}t)|\psi\rangle}$. 
It can be understood as a walk in the Hilbert space, 
where we expect that all states are visited with some probability. 
An important point is that the general form of the transition probability
\beq
P_t(n'|n) \ \ = \ \ |\langle n'|e^{-i\mathcal{H}t}|n\rangle|^2 
\label{tr_prob1}
\eeq
depends on $n$ and $n'$ separately, whereas the spatial transition probability, 
characterizing spatial localization (see App.\ref{app:P}), depends only on the 
difference $|r'-r|$. This is a consequence of the translational invariance in real 
space after disorder averaging. 
It follows that both $\mathcal{P}(t)$ and $\mathcal{N}_{\text{states}}(t)$, 
as well as the asymptotic value $\mathcal{N}_{\infty}$ 
dependent in general on the initial state~$|\psi\rangle$.
Let us re-write \Eq{e10} in a way that emphasize this point:
\beq
\mathcal{N}_{\infty}^{-1} 
\ = \ \lim_{\epsilon\to0} \epsilon 
\int_0^\infty|\langle\psi|e^{-i\mathcal{H}t}|\psi\rangle|^2 e^{-\epsilon t}dt \; .
\label{rprob0}
\eeq
The dependence on the initial state $\psi$ becomes obvious when we choose it to be 
an eigenstate $\ket{E_{\alpha}}$ of $\mathcal{H}$, which gives $\mathcal{N}_{\infty}=1$. 
In this case, the system remains in the initial state for all times. An initial 
state $\ket{n_0}$ within the energy-shell might be similar to an eigenstate of~$\mathcal{H}$. 
Then the question is whether it will visit only a fraction or the entire Hilbert space. 
From the definition \Eq{e10} and the spectral decomposition \Eq{e8} it follows that 
\beq
\mathcal{N}_{\infty}^{-1}
\ = \ \lim_{\epsilon\to0}\epsilon^2\sum_{\alpha,\beta}\frac{p_{\alpha} p_{\beta}}{\epsilon^2+(E_{\alpha}-E_{\beta})^2}  \; .
\eeq
For a non-degenerate discrete spectrum we obtain \Eq{e11}, 
and hence identify  $\mathcal{N}_{\infty}$ as the number of participating eigenstates. 
Recall that $\mathcal{N}_E$ is the effective Hilbert space dimension.
Accordingly, the following ratio can be used as a measure for localization \cite{Heller_87}:  
\beq
\mathcal{F} \ \ = \ \ \frac{\mathcal{N}_{\infty}}{\mathcal{N}_E} \; .
\eeq  
It is the fraction of {\em eigenstates} within the energy shell 
that participate in the dynamics, and by \Eq{e10} it is 
also the fraction of {\em cells} that will be occupied by the 
probability distribution in the long time limit. 
It is implicitly assumed that the preparation~$\psi$
is localized in a small region of the energy shell, 
as in the case of a Gaussian wavepacket in a chaotic billiard. 
The dimension of Hilbert space $\mathcal{N}_E$ is proportional to the volume, 
while $\mathcal{N}_{\infty}$ would be volume-independent 
if the eigenstates were localized. Accordingly a vanishingly 
small $\mathcal{F}$ constitutes an indication for 
a strong quantum localization effect.

In a practical calculation, it is essential to use consistent 
mathematical definitions for $\mathcal{N}_E$ and $\mathcal{N}_{\infty}$.
It is convenient to express them in terms of a the smoothed spectral density, 
the so called local density of states, 
\beq
\varrho_\epsilon(\omega) \ &=& \ \frac{1}{\pi} \mbox{Im}
\BraKet{\psi}{(\omega-i\epsilon-\mathcal{H})^{-1}}{\psi} \\
&=&  \BraKet{\psi}{ \delta(\omega-\mathcal{H})}{\psi} \; .
\label{density00}
\eeq
The survival probability $\mathcal{P}(t)$ is the squared absolute value 
of its Fourier transform, while for the participation number we get 
\beq
\mathcal{N}_{\infty}^{-1}
\ = \ 2\pi \lim_{\epsilon\to0}\epsilon\int[\varrho_\epsilon(\omega)]^2d\omega \; .
\label{r_density}
\eeq
The right-hand-side can be treated within the recursive projection 
method \cite{ziegler03,ziegler11}, see Sec.\ref{s:P}. 
On equal footing, we can define the effective Hilbert space 
dimension from a semiclassical calculation: 
\beq
\mathcal{N}_{E}^{-1}
\ = \ 2\pi \int[\varrho_{\text{cl}}(\omega)]^2d\omega \; .
\eeq
In the absence of a classical limit $\varrho_{\text{cl}}(\omega)$ can be regarded 
as the envelope of $\varrho_\epsilon(\omega)$.

\section{Quantum ergodization}

The simplest dynamical scenario of ergodization concerns a fully-chaotic Sinai billiard. 
The function $\mathcal{N}_{\text{cells}}(t)$ approaches the saturation 
value $\sim \mathcal{N}_E$ after a very short ergodic time~$t_{\text{erg}}$ 
that is determined by the curvature of the walls. 
Similarly, the function $\mathcal{N}_{\text{states}}(t)$ approaches the asymptotic 
value $\mathcal{N}_{\infty} = \mathcal{N}_E$ in the classical case, 
or $\mathcal{N}_{\infty} = \mathcal{N}_E/3$ in the quantum case. 
The factor ${\mathcal{F}_{\text{RMT}}=1/3}$ reflects the Gaussian statistics of a 
random-wave amplitude. 
Not all possible preparations look like random waves.
If we start a wavepacket at the vicinity of an unstable 
periodic orbit,  $\mathcal{F}$ might be further suppressed 
due to recurrences. The suppression factor depends 
on the Lyapunov instability exponent~$\lambda$, and is given
by a variation of a formula of the following type \cite{scar1,scar2}:    
\beq \label{e20}
\mathcal{F} 
\ \ \approx \ \  
\left[ \sum_{s=-\infty}^{\infty} \frac{1}{\cosh(\lambda s)} \right]^{-1} \mathcal{F}_{\text{RMT}} \; .
\eeq
The study of a periodically kicked Bosonic Joshephson Junction provides 
a nice demonstration for this formula \cite{ckt}. 
One should regard ``scarring" as a {\em weak localization} effect.

Let us describe how $\mathcal{N}_{\text{states}}(t)$ 
looks like in the absence of strong localization effect.
We still have in mind the simplest dynamical scenario 
of ergodization that concerns a fully-chaotic Sinai billiard.  
This function differs from $\mathcal{N}_{\text{cells}}(t)$ 
because the time-averaged~$\rho(t)$ gives a large weight 
to the initial preparation, which diminishes only 
algebraically (as~$1/t$) in the ${t\rightarrow\infty}$ limit.
To make this point clear, notice that the survival probability $\mathcal{P}(t)$ 
decays within a time $\tau_E$. For short times it is roughly 
described by recurrences-free quantum exploration function  
\beq
\mathcal{N}^{(0)}_{\text{states}}(t) \ \ = \ \ \frac{t}{\tau_E} \; .
\eeq 
After that, there are recurrences whose long time 
average saturates asymptotically to the value $1/\mathcal{N}_{\infty}$. 
Therefore we get roughly 
\beq \label{e13}
\mathcal{N}_{\text{states}}(t) \ \ \approx \ \ 
\left[ \frac{1}{t/\tau_E} + \frac{1}{\mathcal{N}_{\infty}} \right]^{-1} \; .
\eeq   
This expression is characterized by a crossover time that we call ``quantum breaktime". 
In the present example we identify the quantum breaktime 
with the Heisenberg time, namely, 
\beq
t^* [\text{for ergodic system}] \ \ \sim \ \ \mathcal{N}_{\infty} \tau_E \ \ \sim \ \ t_H
\ .
\eeq

We conclude that in the simple quantum ergodization scenario 
the two functions  $\mathcal{N}_{\text{cells}}(t)$ 
and  $\mathcal{N}_{\text{states}}(t)$ involve completely different time scales.
The former  get ergodized after a short classical time~$t_{\text{erg}}$, 
while the latter saturates only after the quantum Heisenberg time~$t_H$. 
This picture changes completely if we have a strong localization effect, 
as discussed in the next section.

In more complicated systems, there are two types of modifications that are 
expected in \Eq{e13}. Weak localization corrections (scar-theory related) 
can be deduced from the study of the survival probability. Recall that 
$\mathcal{P}(t)$ is determined by the Fourier transform of the local density of 
states $\varrho(\omega)$. The short time behavior reflects the envelope of 
this spectral function, while for longer times a power law decay $\mathcal{P}(t) \sim t^{-\gamma}$ 
would reflect the fractal dimension of the participating eigenstates \cite{lea1,lea2}. 
Accordingly,  
\beq
\mathcal{N}_{\text{states}}(t) \ \ \sim \ \ 
\left[ \frac{\text{const}}{t^{\gamma}} + \frac{1}{\mathcal{N}_{\infty}} \right]^{-1} \; .
\eeq
By definition, the asymptotic value $\mathcal{N}_{\infty}$ reflects the number 
of participating eigenstates, and can be deduced semiclassically too.
This will be discussed in the following sections. We refer to the possibility of 
having ${\mathcal{N}_{\infty} \ll \mathcal{N}_E}$ as a strong localization effect.

\begin{figure}
\begin{center}
\includegraphics[width=7.5cm]{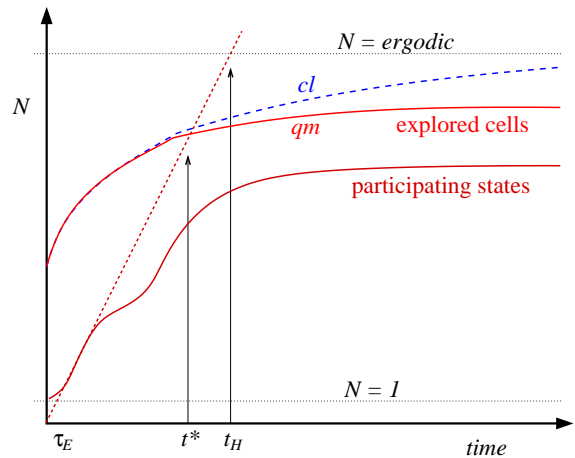}

\caption{
\rmrk{Schematic description for the exploration of phase-space versus time. 
For example it might be the phase-space of billiard that consists 
of connected-boxes (see text). Diffusion-like dynamics is assumed.}
The participation number is bounded between the minimal value $\mathcal{N}=1$ and the maximal ergodic value $\mathcal{N}_E$, 
which are represented by horizontal dotted lines.  The upper dashed and solid lines 
are the Boltzmann exploration function $\mathcal{N}_{\text{cells}}(t)$ 
in the classical and in the quantum evolution, respectively. 
They depart at the breaktime $t^*$, and consequently a quantum localization effect manifests itself. 
The lower solid and dashed lines are $\mathcal{N}_{\text{states}}(t)$  
and its upper bound  $\mathcal{N}_{\text{states}}^{(0)}(t)$, respectively. 
The former reflects the survival probability, while the latter is determined 
by the width of the energy-shell and intersects $\mathcal{N}_E$ at the Heisenberg time~$t_H$.
The actual breaktime $t^*$ is determined by the intersection with $\mathcal{N}_{\text{cells}}(t)$.
Quantum localization is implied if $t^*<t_H$.  
\rmrk{If the dynamics were fully chaotic as for Sinai billiard,}   
then $\mathcal{N}_{\text{cells}}(t)$  would reach the ergodic limit almost immediately, 
and consequently we would have $t^* \sim t_H$ implying no localization. 
In the absence of recurrences $\mathcal{N}_{\text{states}}(t) \sim \mathcal{N}_{\text{states}}^{(0)}(t)$
implying full quantum-ergodization without even a weak localization effect. 
}
\label{fig:NvsT}
\end{center}
\end{figure}

\section{Strong localization}

\rmrk{Strong localization means that the breaktime $t^*$ is shorter
than the Heisenberg time~$t_H$. The prototype example is 
of course Anderson localization where the breaktime is related 
to the strength of the disorder and not to the total volume of the system, 
leading to a finite localization length in space.
Here we would like to formulate the notion of ``strong localization" 
in a more general way, within the HSL phase-space framework.}

If we have a strong quantum-localization effect the role 
of~$t_{\text{erg}}$ and~$t_H$ is taken by the quantum breaktime~$t^*$, 
as illustrated in \Fig{fig:NvsT}. \rmrk{For presentation purpose  
we still consider a variant of the Anderson model: 
an array of connected chaotic boxes, with a particle that can migrate from box to box via small holes.} 
In such a scenario, the classical exploration 
of phase-space is slow, diffusive-like. This means that $t_{\text{erg}}$ is very 
large, and can be defined as the time that it takes to diffuse over 
the whole volume of the array.  The basic conjecture is that quantum-to-classical 
correspondence (QCC) is maintained as long as~$t$ is smaller than the {\em running} 
Heisenberg time. The latter refers, by definition, not to the total volume but 
to the {\em explored} volume. Accordingly, the necessary condition 
for QCC is ${t \ll [\mathcal{N}_{\text{cells}}(t)/\mathcal{N}_E] t_H}$. 
This can be re-written as ${\mathcal{N}^{(0)}_{\text{states}}(t) \ll \mathcal{N}_{\text{cells}}(t)}$.
Using a classical estimate for the number of explored cells, 
we deduce a more practical version for this condition:  
\beq \label{e16}
\mathcal{N}^{(0)}_{\text{states}}(t) \ \ll \ \mathcal{N}^{cl}_{\text{cells}}(t), 
\ \ \ \ \ \ \ \ \text{[QCC condition]}  \; .
\eeq  
Clearly, for diffusive-like dynamics, the breakdown of this inequality 
might happen before the ergodic time as illustrated in \Fig{fig:NvsT}. 
Under such circumstances strong quantum localization is expected.

\rmrk{The reasoning above is not rigorous, still it provides the correct predictions 
as far as Anderson localization is concerned, and we believe that it can be trusted 
in more general circumstances (see below).} 
It offers an illuminating {\em semi-classical alternative to the formal Anderson criterion} \cite{anderson58}. 
Note that the use of the  Anderson criterion is restricted to disordered lattices that have well defined ``connectivity", 
while semiclassics can take into account the implications of complex phase-space structures.

The semiclassical localization criterion determines whether strong localization 
effect is expected, and provides a practical estimate for the localization volume. 
The procedure is simple: Given a dynamical system we have to find $\mathcal{N}^{cl}_{\text{cells}}(t)$. 
Then we estimate $t^*$ as the time when \Eq{e16} breaks down, and the localization 
volume as $\mathcal{N}^{cl}_{\text{cells}}(t^*)$. 
We can further conjecture that the saturation value $\mathcal{N}_{\infty}$  
is of the same order of magnitude (``one parameter scaling").
The implicit assumption in this procedure is that the 
localization effect correlates the functions $\mathcal{N}_{\text{cells}}(t)$
and $\mathcal{N}_{\text{states}}(t)$, as illustrated in \Fig{fig:NvsT}.

In a diffusive system, the classical exploration grows asymptotically 
like~$t^{1/2}$ in 1D, like~$t/\ln(t)$ in 2D, and like~$t$, 
with small corrections, for higher dimensions \cite{brk}. 
Schematically, we write 
\beq
\mathcal{N}^{cl}_{\text{cells}}(t) \ \sim \ \text{const} \ 
+ \ g \times \left(\frac{t}{\tau_E}\right)^{\alpha} \; .
\eeq
From \Eq{e16} we deduce that for sub-linear time dependence (${\alpha<1}$), 
we always have quantum localization, as in the case of diffusion in 1D/2D.  
On the other hand, if the asymptotic rate of exploration is linear (${\alpha=1}$), 
then the prefactor~$g$ becomes important: the condition for quantum localization 
becomes ${g<1}$, implying a mobility edge. 
For a diffusive particle in a $d$-dimensional array of connected-boxes, 
one can easily show that ${g=(k\ell)^{d-1}(\ell/L)}$, 
where $k$ is the wavenumber, $\ell$ is the linear size of each box, 
and $L$ is the mean free path for box-to-box migration.

More generally, in quantized chaotic systems, the exploration of phase 
space can be slowed down by cantori (remnants of Kolmogorov-Arnold-Moser tori) or due to the sparsity 
of the Arnold web. Turning to the many body localization problem, a mobility 
edge is implied in any dimension, provided the classical rate of exploration 
depends on energy, such that above some $E_c$ the exploration is linear in time 
and fast enough. For any system with more than two degrees of freedom, the resonances 
between the coupled degrees of freedom form an ``Arnold web" in phase-space.
As energy is increased, the competition is between the width of the filaments 
and their density. Typically, above some critical energy the filaments overlap
and we get fast exploration -- this is called Chirikov criterion. But also, 
below this critical energy there is spreading which is called ``Arnold diffusion". 
Thus one expects in the latter case quantum localization too \cite{ArQ1,ArQ2}.

An interesting  application of the above framework is in order to determine 
the criterion for superfluidity in atomtronic circuits. It has been 
demonstrated numerically \cite{sfc} that superfluidity can be dynamically stable 
in rings that are described by the Bose-Hubbard Hamiltonian. The explanation 
of this stability requires ``quantum localization" in a region where the Arnold 
diffusion prevails.

\section{The recursive projection method}
\label{s:P}

In principle, the participation number $\mathcal{N}_{\infty}$ can be determined 
numerically, provided the total dimension of Hilbert space is not too large.  
But if we want analytical results, we have to adopt an appropriate method. 
One possibility is to use semiclassics. The leading order semiclassics is merely 
the calculation of phase space-volumes, and hence can provide us $\mathcal{N}_E$. 
But then we have to adopt higher order semiclassics 
(periodic orbit theory or RMT phenomenology) in order to say anything 
regarding~$\mathcal{F}$. If we want to get results for strong localization, we have 
to use complicated summation methods, or to be satisfied with the breaktime 
phenomenology.      

In the next section, we would like to demonstrate an {\em exact} 
calculation of $\mathcal{N}_{\infty}$ using the recursive projection method (RPM).
Given an initial state $|\psi\rangle$, this method facilitates 
the calculation of the return amplitude. A Laplace transformation ${t\to z}$ 
maps the evolution operator to the resolvent $(z-H)^{-1}$, while the inverse mapping 
is provided by a Cauchy integration on the complex $z$-plane. The advantage 
of using the resolvent is its presentation as a rational function with 
polynomials $Q_{N}(z)$ and $P_{N-1}(z)$:
\beq
G_0(z) \ = \ \langle\psi|(z-H)^{-1}|\psi\rangle=\frac{P_{N-1}(z)}{Q_{N}(z)} \ .
\label{rational}
\eeq
Its poles of $G_0(z)$ are determined by the the zeros of $Q_{N}(z)$, 
and they are the eigenvalues of~$H$. The RPM provides an efficient procedure 
to calculate these polynomials. Starting from the fact that the resolvent 
can be expanded in powers of $H$, the resulting geometric series can be 
understood as a random walk in Hilbert space, where each sub-Hilbert space 
is visited an arbitrary number of times. In general, this series, also known 
as the Neumann series of the resolvent, has an infinite number of terms and 
is valid only within its radius of convergence. The main idea of the RPM is 
to re-organize the geometric series as a directed random walk through the 
$N$-dimensional Hilbert space, where each sub-Hilbert space is visited only 
once \cite{ziegler03,ziegler11}, rather than an arbitrary number of times. 

Following the recipe described in Ref. \cite{ziegler11}, we obtain the 
rational function \Eq{rational} in $N$ calculational steps. Since this 
still gives quite lengthy expressions for the polynomials with typical values 
$N=30...50$, we plot only the resulting values of the return probability as 
a function of time and $\mathcal{N}_{\infty}$ as a function of $N$ here.

\section{Bosonic Josephson Junction}

Small closed systems have the advantage that: (i) they can be realized 
experimentally and (ii) theoretical calculations are simple and in many 
cases can be performed exactly for finite dimension $N$. Prominent examples 
are the Jaynes-Cummings model \cite{jaynes63,cummings65}, which describes the 
interaction of a two-level system in an optical cavity (cf. App.\ref{app:jcm}), 
coupled photon cavities \cite{cqed}, and Josephson tunneling junctions for 
bosonic atoms in a double well potential
\cite{milburn97,ketterle04,oberthaler05,Boukobza,gati06,oberthaler07,cohen10}.
Interacting bosons in a double-well potential are described in 
a two-mode approximation by the Bose-Hubbard Hamiltonian \cite{oberthaler07}:
\beq
\mathcal{H}_{\text{BH}} = \frac{U}{2}\sum_{j=1}^{2} 
a_{j}^{\dag} a_{j}^{\dag} a_{j} a_{j} 
- \frac{J}{2} \left( a_{2}^{\dag} a_{1} + \text{h.c.} 
\right)
\ ,
\label{ham00}
\eeq
where $a_j$ and $a_j^\dagger$ are the bosonic annihilation and creation 
operators of the two wells, respectively. This Hamiltonian acts on the space 
that is spanned by the Fock basis $\{|N-k,k\rangle\}_{0\le k\le N}$, where 
\beq
\ket{n} \ \ \equiv \ \  |N-n,n\rangle \ \ \equiv \ \ |N-n\rangle\otimes|n\rangle
\eeq
is the state with $N-n$ bosons in the left well and $n$ bosons in the right well. 
The first term of the Hamiltonian describes a local interaction of the bosons, 
enforcing a symmetric distribution in the double well for $U>0$, while the second 
term of $\mathcal{H}_{\text{BH}}$ describes tunneling of bosons between the wells. 
As a characteristic dimensionless parameter of the Bose-Hubbard Hamiltonian we employ 
\beq
u \ \ = \ \ \frac{NU}{J} \ ,
\eeq
which is the ratio of the interaction energy over the tunneling energy.
The factor $N$ takes care of the fact that the interaction energy grows 
like $N^2$ in our model, while the tunneling energy grows like $N$.
    
\subsection{Limiting cases}

We first consider two limiting cases for the bosonic Josephson junction: 
{\bf (i)} the $J{=}0$ case;  {\bf (ii)} the $U{=}0$ case. 
The latter describes e.g. photons in two coupled cavities.
The transition probability in case (i) is
\beq
|\langle N-n,n|e^{-iHt}|N-n,n\rangle|^2 \ = \ 1 \; ,
\eeq
which implies $\mathcal{N}_{\infty}=1$ and describes a trivial case of HSL. 
For case (ii), a simple calculation reveals for the return amplitude \cite{cqed,ziegler11}
\beq
\langle N, 0|e^{-iHt}|N, 0\rangle \ = \ \cos^N(Jt/2)
\label{non_int_exp} \; , 
\label{ret_prob2}
\eeq
when the initial state has all $N$ bosons in one well.
Thus the evolution in the Fock state is periodic with period $2\pi/J$.
For $N=1$, we get $\mathcal{N}_{\infty}=2$, while for $N\gg1$, using the Stirling formula 
(see App.\ref{app:double}), we obtain
\beq
\mathcal{N}_{\infty}\sim\sqrt{\pi N/2} \ ,
\label{scaling0}
\eeq
which clearly indicates the absence of HSL, as we would have anticipated for 
independent bosons.

The most interesting question is what happens if $J,U\ne 0$; i.e., when 
tunneling and interaction compete. This can be addressed either in a 
mean-field approximation, in a semiclassical calculation, or by an exact 
solution of the quantum dynamics, using the recursive projection method 
for $\mathcal{N}_{\infty}$ in \Eq{r_density}. 

\subsection{Pendulum analogy}
\label{sect:pendulum}

It is natural to consider first the simplest possible approximation.
The prototype one-degree of freedom system is the {\em pendulum}. 
The Josephson junction is described formally 
by the same Hamiltonian with conjugate action-angle canonical coordinates 
${(\varphi,n)}$, and with $\hbar \rightarrow (E_C/E_J)^{1/2}$. 
The Bosonic Josephson junction ("dimer") is a different version 
of the pendulum Hamiltonian with $\hbar \rightarrow 1/N$, 
where $N$ is the number of particles. 
Our calculation below refers to the latter, therefore we highlight 
the $N$ dependence of the results. 

Let us summarize a few known results that concern the dimer \cite{ckt}:  
{\bf (1)} For a ground-state preparation $\mathcal{N}_{\infty}$ is of 
order unity. {\bf (2)} For a preparation at the vicinity of the hyperbolic 
classically-unstable point $\mathcal{N}_{\infty} \sim \log(N)$. 
{\bf (3)} For a generic preparation $\mathcal{N}_{\infty} \sim N^{1/2}$. 
The latter is implied by the uncertainty relation $\Delta \varphi \Delta n \sim \hbar \rightarrow 1/N$.  
{\bf (4)} For a periodically driven chaotic dimer $\mathcal{N}_{\infty} \sim N$, 
with scar-theory correction \Eq{e20} in the vicinity of unstable periodic periods.  

\subsection{Mean-field dynamics}
\label{sect:mfa}

\begin{figure}
\begin{center}
\includegraphics[width=7.5cm]{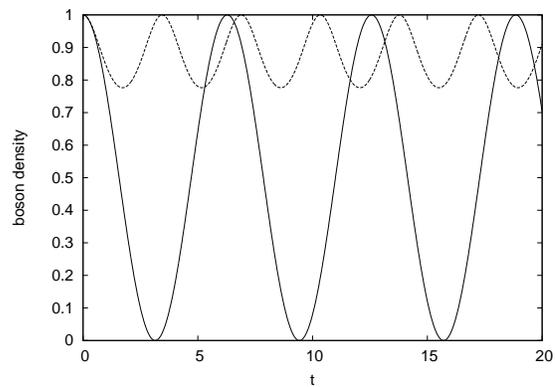}
\caption{
Mean field dynamics of the boson density $N_1(t)/N$ in the left well, 
as described in \Eq{mfa1}. The function $[1+\mbox{cn}(t|y)]/2$ is plotted for 
different values of $y$, namely $y=0.8$ (full curve) and $y=1.1$ (dashed curve). 
\rmrk{Axes units are dimensionless.}
}
\label{fig:jacobi}
\end{center}
\end{figure}

We consider the initial preparation ${\ket{N,0}}$. Then the expectation 
values for the number of bosons in the left well at time $t$ is  
\beq
N_1(t) &=& \langle\psi(t)|a_1^\dagger a_1|\psi(t)\rangle \\ 
&=& \langle\psi(0)|e^{iHt}a_1^\dagger a_1 e^{-iHt}|\psi(0)\rangle \ .
\eeq
For this quantity, a mean field approximation exists in the form of a
nonlinear Schr\"{o}dinger (Gross-Pitaevskii) equation \cite{milburn97,eilbeck85}.
The solution reads
\beq
\label{mfa1}
N_1(t) &=& \frac{N}{2}\left[1+ \mbox{cn}(2J t|N^2/N_c^2)\right] , \\ 
N_c &=& \frac{2N}{u}=\frac{2J}{U} 
\eeq
with the initial value $N_1(0)=N$. The Jacobian elliptic function 
$\mbox{cn}(x|y)$ \cite{abramowitz} is periodic in the first argument for $y\ne1$ 
and changes its behavior qualitatively when the second argument $y=(u/2)^2$ crosses 
over from $y<1$ to $y>1$. The corresponding spectrum is equidistant, with 
$E_k=(2k+1)\pi J/K(u^2/4)$, where $K(m)$ is the integral
\[
K(m)=\int_0^{\pi/2}\frac{1}{\sqrt{1-m\sin^2\phi}}d\phi
\ .
\]
Two examples for $N_1(t)$ are depicted in \Fig{fig:jacobi}. In 
terms of our bosonic double well system, the behavior for ${N<N_c}$ describes 
unhindered tunneling of the bosons from one well to the other, whereas 
for ${N>N_c}$, only a fraction of the bosons are allowed to tunnel. This 
behavior is known as self-trapping \cite{milburn97,eilbeck85} or mode-locking 
\cite{YYB_97,YYB_00,YYB_02} and reminds us of HSL, since it indicates that 
the bosons are trapped in the well from which the particle dynamics started 
originally, and it cannot explore the entire Hilbert space spanned by $\{ \ket{n} \}$.

\subsection{Semiclassical dynamics}

The two-site Bose-Hubbard Hamiltonian for $N$ bosons is equivalent to 
a nonlinear $SU(2)$ spin model in $N+1$ dimensions (cf. App.\ref{app:spin})
\beq
H_{S} \ = \ UL_z^2 - JL_x
\ .
\label{spin_ham}
\eeq 
Thus the semiclassical dynamics can be represented in a spherical phase-space. 
The expectation values of ${(L_x,L_y,L_z)}$ are known as the Bloch vector.
In order to visualize a wavefunction, the Bloch vector is not sufficient: we have 
to think about the Wigner function $\rho_W(\theta,\varphi)$. For coherent-state 
preparation, the Wigner function looks like a minimal Gaussian. In the leading 
order semiclassics, so called truncated Wigner approximation, the  Wigner ``distribution" 
propagates by the classical equations of motion, that are formally identical to the 
mean field equations. Note however that the semiclassical perspective is beyond mean field 
because we propagate a {\em cloud} and not a single point on the Bloch sphere. 

In the following, we focus on two initial states with different 
energies, namely $|N,0\rangle$ and $|N/2,N/2\rangle$, assuming that $N$ is 
even. For $J=0$ these are eigenstates of $\mathcal{H}_{\text{BH}}$ 
with energies $E=U N^2/2$ and $E=U N^2/4$, respectively. 
On the Bloch sphere, these states are represented by ``distributions" 
that reside in the North-pole and along the the Equator respectively.

\subsection{Semiclassical $\mathcal{N}_E$}

The semiclassical perspective can provide us a qualitative expectation 
regarding the dependence of  $\mathcal{N}_{\infty}$ on~$u$.
In leading order, we expect ${\mathcal{N}_{\infty} \sim \mathcal{N}_E}$, 
which is simply the phase-space volume that is filled by the evolving
cloud. We recall \cite{cohen10} that for $u>1$ a separatrix appears 
with hyperbolic point at ${(\theta, \varphi) = (\pi/2,\pi)}$ on the Bloch sphere. 
As $u$ is increased, this separatrix stretches further along 
the $\theta=\pi/2$ axis. For $u>2$ it goes beyond the North Pole.
This is the reason why a North Pole preparation becomes self-trapped: 
it is locked inside the North island, and cannot migrate to the 
South island (assuming that tunneling is ignored). 
From this picture it follows that $\mathcal{N}_{\infty}$ 
is expected to drop sharply for ${u>2}$.

A somewhat different $u$ dependence is expected for an Equator state preparation.
Here part of the cloud is trapped at the vicinity of the hyperbolic point 
as soon as the separatrix is created (${u>1}$). The drop in $\mathcal{N}_{\infty}$ 
is expected to be less dramatic because only a small portion of the cloud is affected. 

The total Hilbert space dimension is ${(N{+}1)}$, which is the number of Planck-cells 
on the spherical phase space. The number of participating cells $\mathcal{N}_E$, 
within the energy shell, can be found analytically in terms of a phase-space integral.    
The scaling of  $\mathcal{N}_E$ with respect to ${(N{+}1)}$ depends on the initial 
preparation: For the $|N,0\rangle$ North-Pole preparation 
we expect ${\mathcal{N}_E \propto (N{+}1)^{1/2}}$,  
reflecting the area of a minimal wave-packet; 
while for the $|N/2,N/2\rangle$ preparation 
we expect ${\mathcal{N}_E \propto (N{+}1)}$, 
reflecting that the number of energy-contours that 
intersect the Equator scales linearly with the total 
Hilbert space dimension.   
For a fixed finite~$N$ the $u$~dependence of $\mathcal{N}_E$ 
might be described by some scaling function~$f(u)$. 
We shall find this function using an exact RPM quantum calculation.

\begin{figure}

\begin{center}
\includegraphics[width=7.5cm]{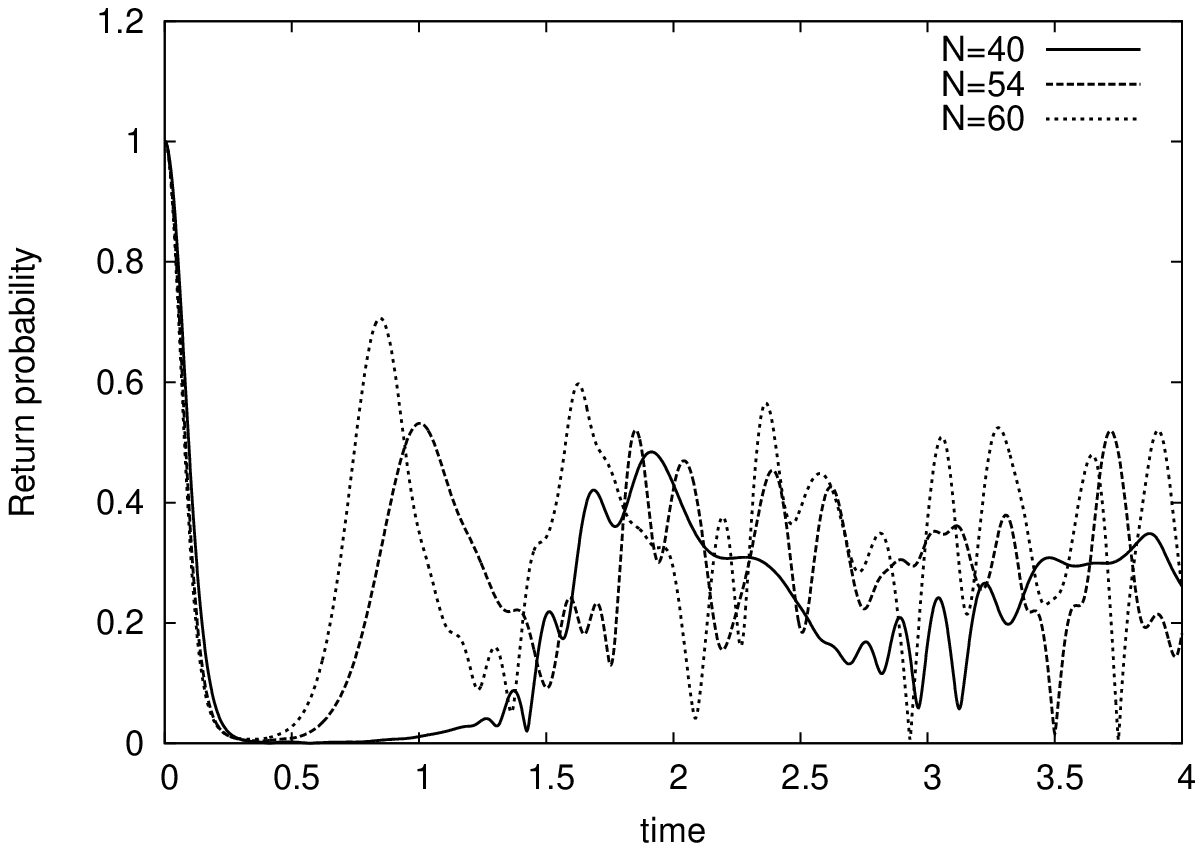}
\caption{
Time-dependent return probability $\mathcal{P}(t)$ for the initial state $\ket{N,0}$ 
with $N=40, 54, 60$. 
The model parameters are $U=0.16$ and $J=4$ with the dimensionless interaction $u=N/25$. 
The critical value $u_c=2$ for mean field self-trapping is $N_c=50$.
\rmrk{Axes units are dimensionless.}
}
\label{fig:dyn1}
\end{center}

\begin{center}
\includegraphics[width=7.5cm]{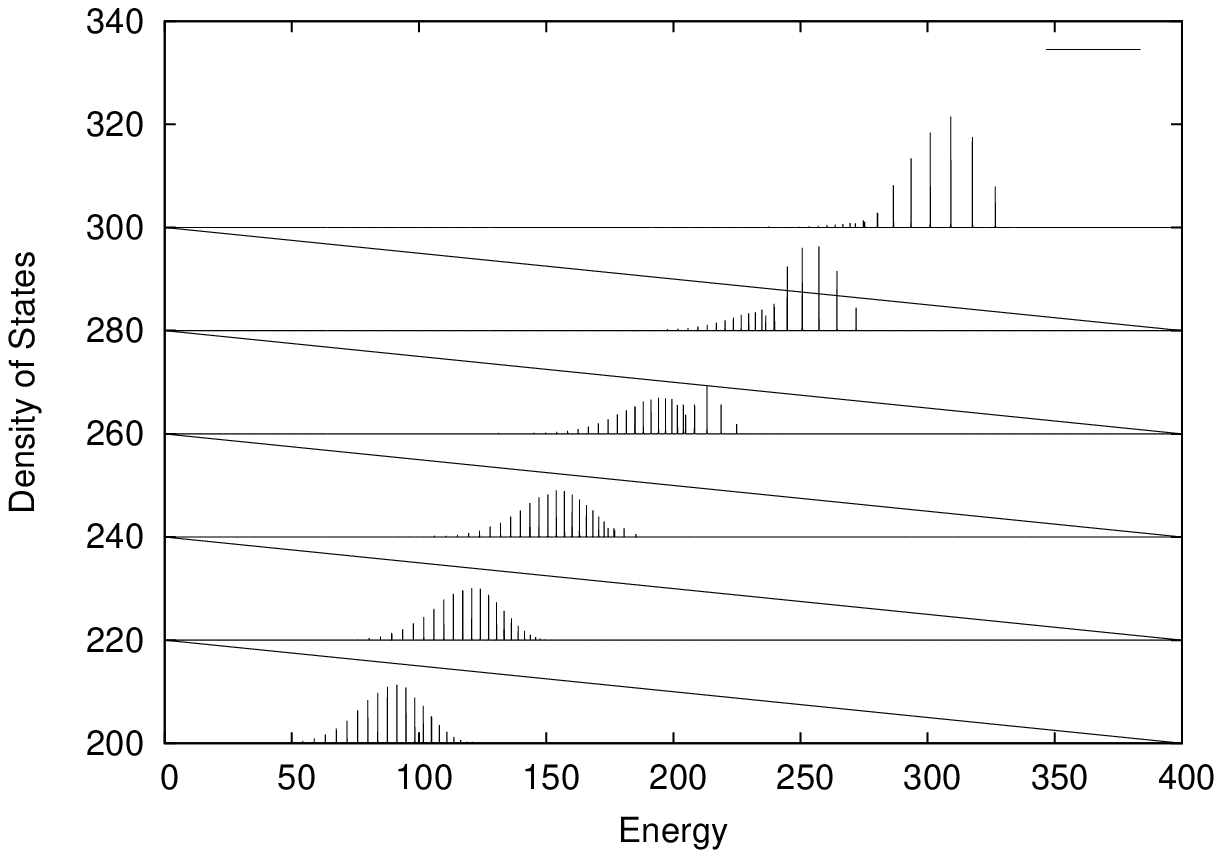}
\caption{ 
The spectral density $\varrho(\omega)$ with respect to the state $|N,0\rangle$, 
for several values of $u$ between $u=1$ and $u=2.6$ (from bottom to top). 
Starting at ${u \approx 1.8}$ the spectral density becomes fragmented, 
see text for details. 
\rmrk{Axes units are dimensionless.}
}
\label{fig:dos}
\end{center}

\end{figure}

\subsection{Exact calculation - $\mathcal{P}(t)$}

In \Fig{fig:dyn1}, we plot the time-dependent return probability $\mathcal{P}(t)$ 
for the initial state $\ket{0,N}$. It is obtained analytically from the resolvent 
by a Cauchy integration. We see that on short time scales the overlap 
with the initial state decays very quickly with the same behavior for different values of~$N$, 
but the recurrences are somewhat different for $N<N_c$ and $N>N_c$. 
This agrees with the short time behavior of the mean field solution in \Fig{fig:jacobi}. 
On the other hand, the dynamical behavior of the return probability does not clearly 
distinguish two qualitatively different regimes, unlike the mean field dynamics. 
Namely, the periodic behavior with a clear signature of self-trapping, 
which is visible on a relative short time scale of the mean-field dynamics, 
is not manifest in the quantum dynamics due to its rather irregular behavior. 

The somewhat erratic time dependence of $\mathcal{P}(t)$ is related 
to a qualitative change of the spectral density $\varrho(\omega)$, 
as depicted in \Fig{fig:dos}: 
As~$u$ is increased, and the critical point ${u_c \sim 2}$ is approached,  
there is a characteristic fragmentation of the spectrum into  
non-degenerate low-energy region and doubly-degenerate high energy region.  
This is quite different from the equidistant energy levels of the
mean field approximation in Sect.\ref{sect:mfa}.

\begin{figure}
\begin{center}
\includegraphics[width=7.5cm]{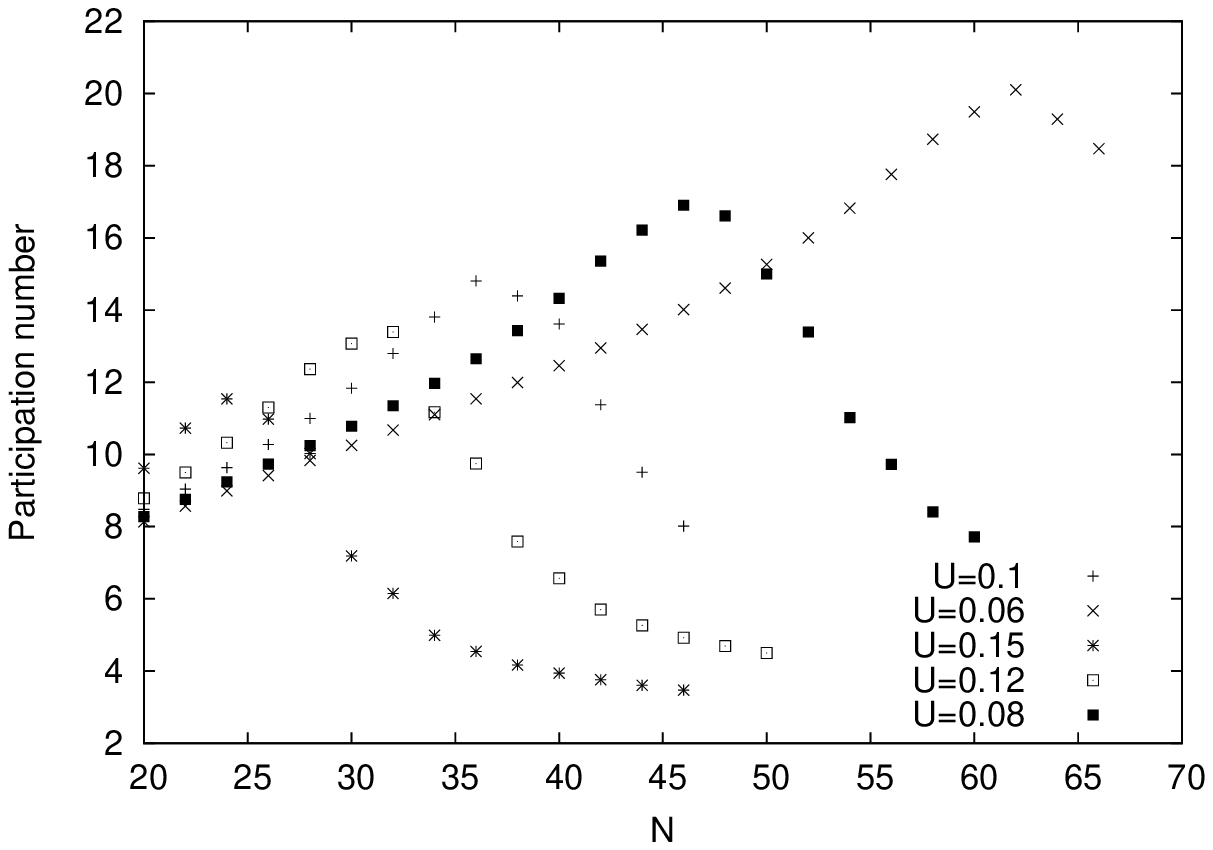}
\includegraphics[width=7.5cm]{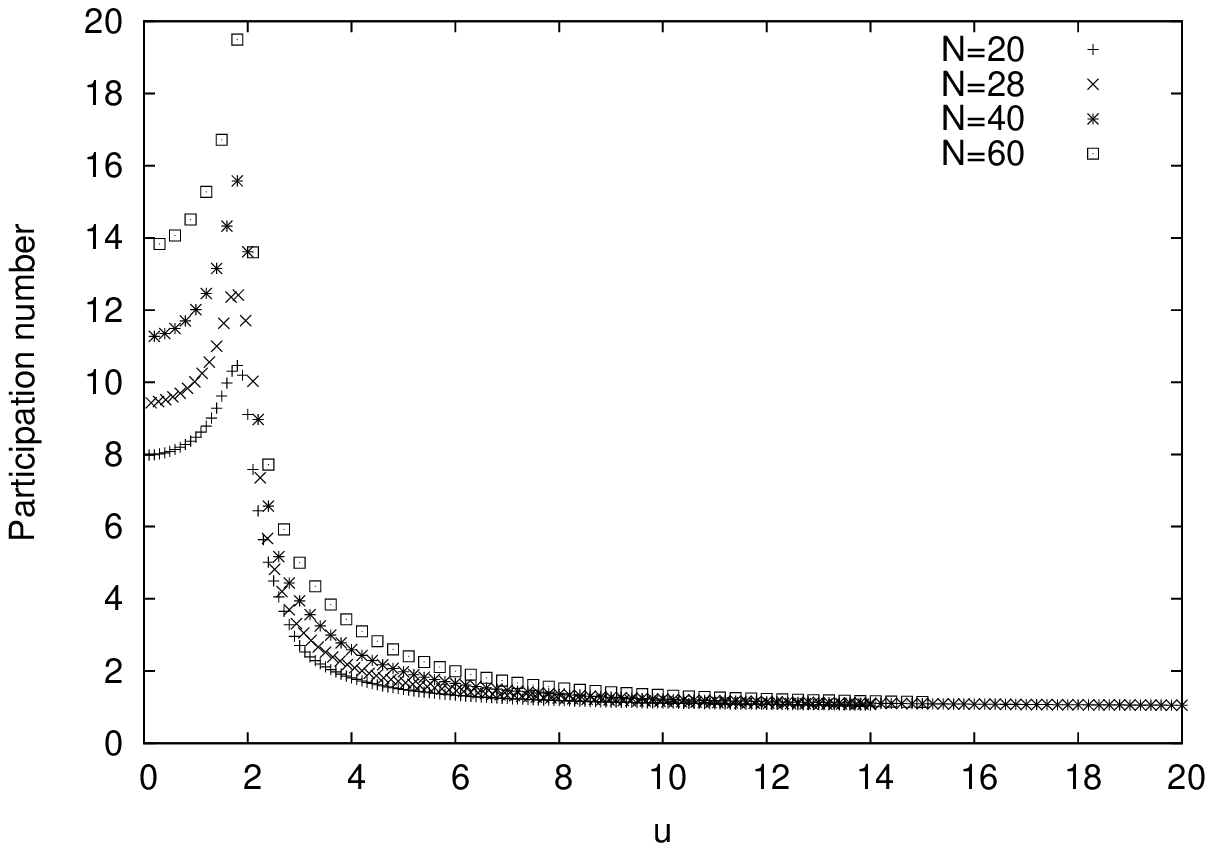}
\caption{
The participation number $\mathcal{N}_{\infty}$ for the initial state $|N,0\rangle$ and with $J=2$ 
as a function of the boson number $N$ for different values of the interaction $U$ (upper panel)
and as a function of $u$ (lower panel).
}
\label{fig:loc0}
\end{center}

\begin{center}
\includegraphics[width=7.5cm]{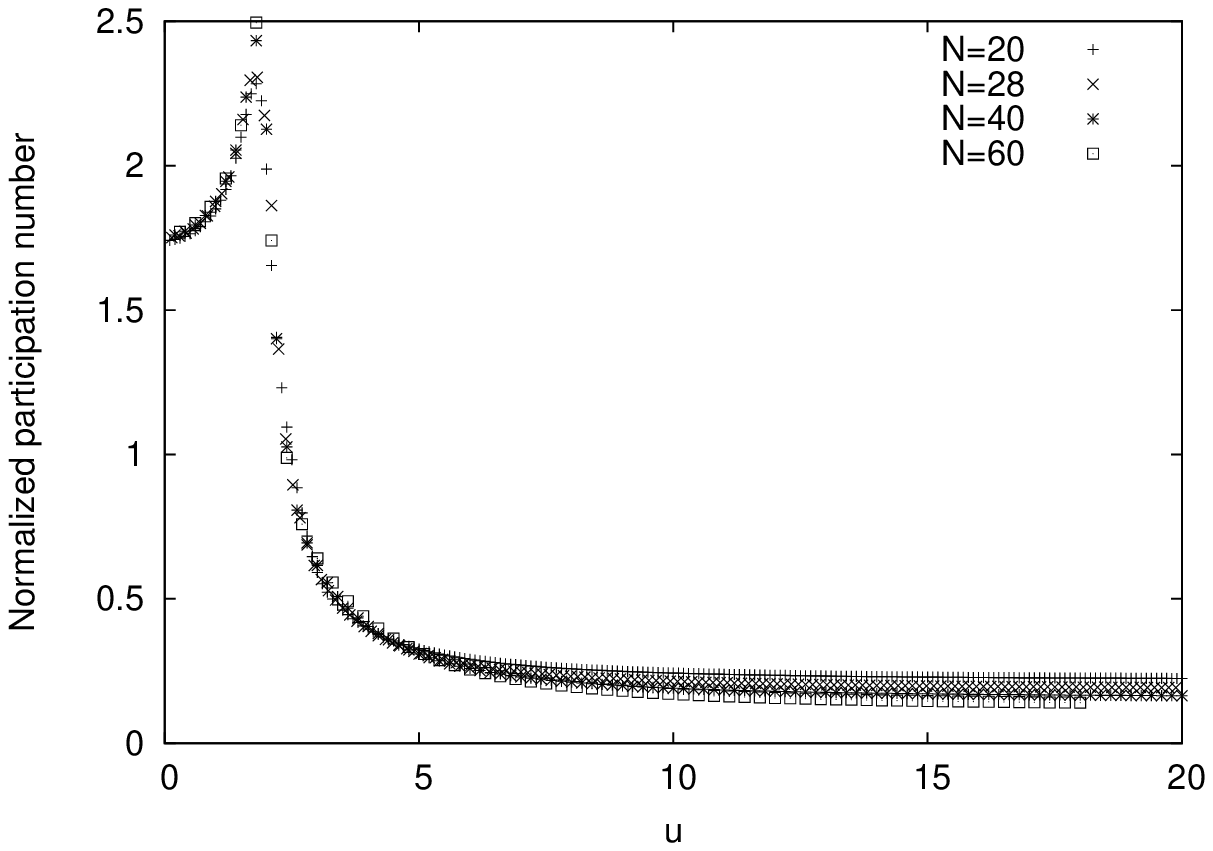}
\caption{
The normalized participation number $\mathcal{N}_{\infty}/\sqrt{N+1}$ for the initial state $|N,0\rangle$ and with $J=2$ 
as a function of $u$. \rmrk{Axes units are dimensionless.}
}
\label{fig:pn_b}
\end{center}
\end{figure}

\begin{figure}
\begin{center}
\includegraphics[width=7.5cm]{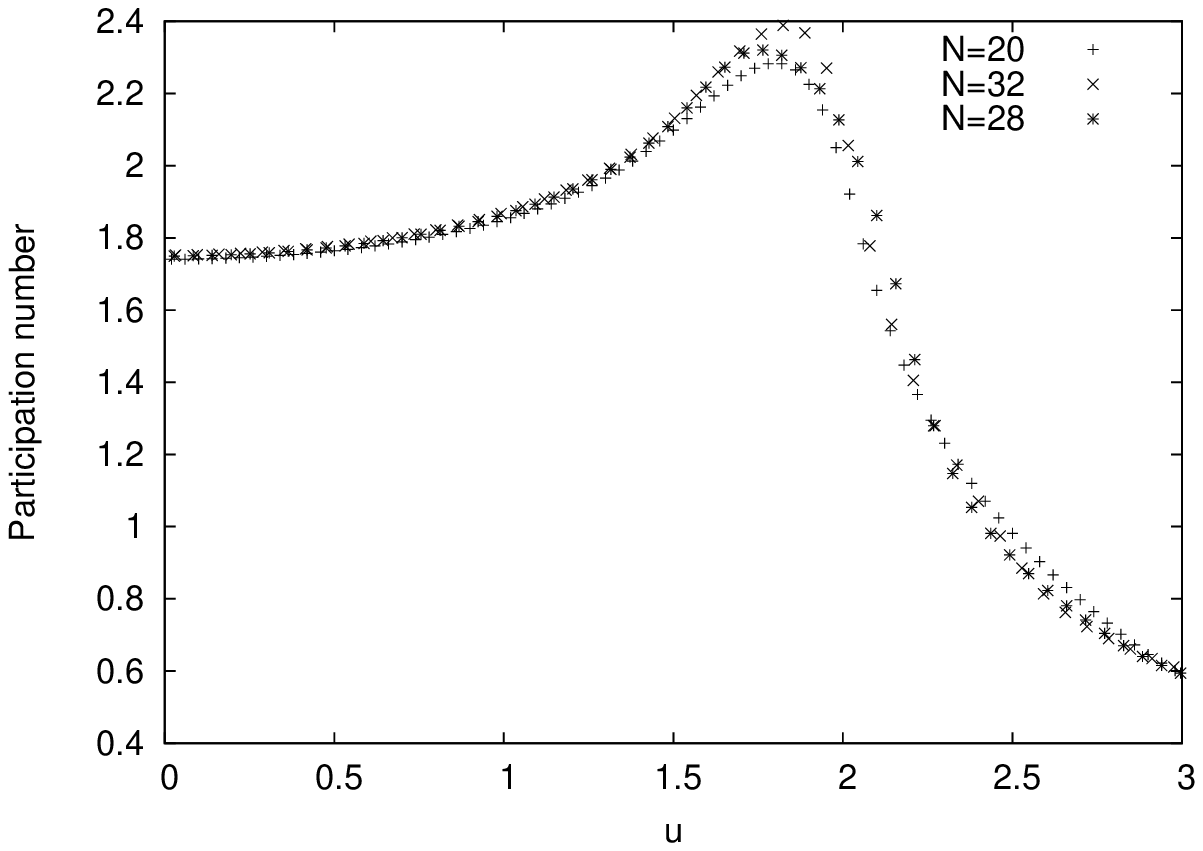}
\includegraphics[width=7.5cm]{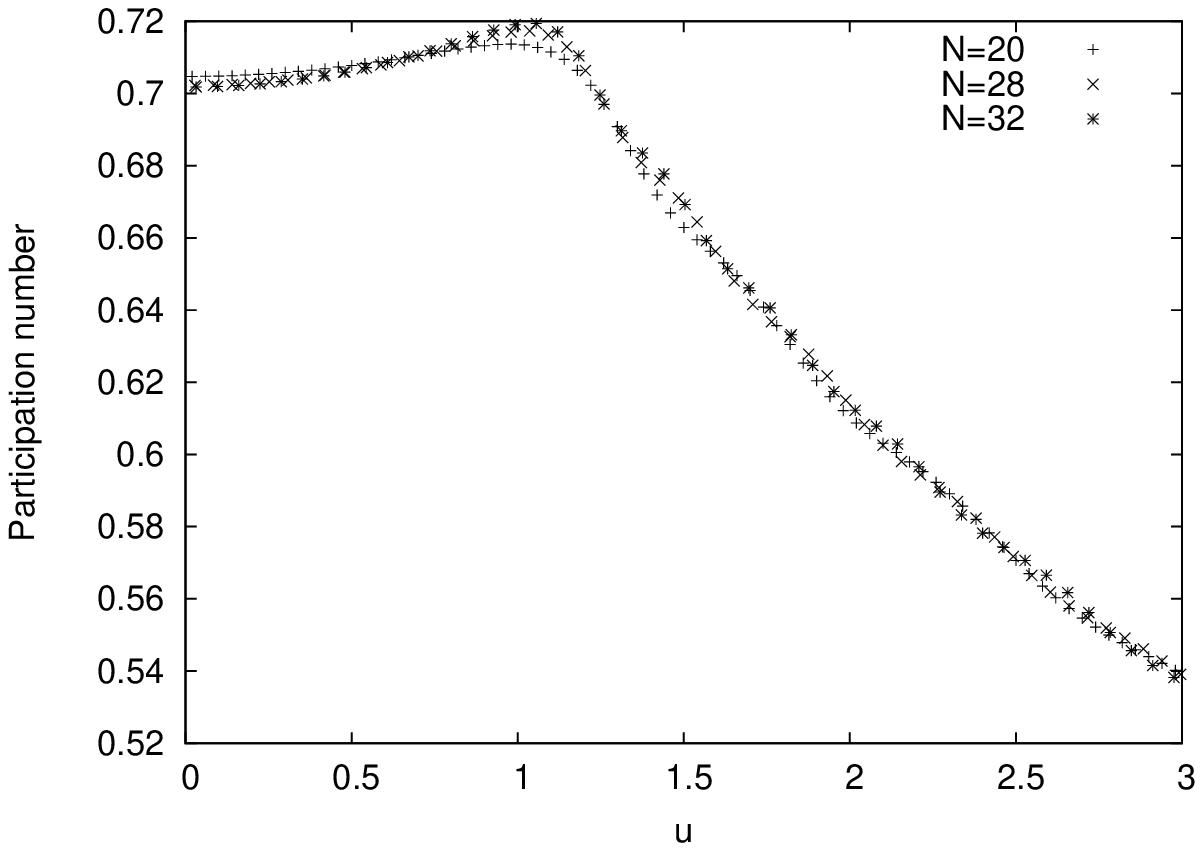}
\caption{
The normalized participation number $\mathcal{N}_{\infty}/(N{+}1)^\alpha$
as a function of the parameter $u$ for different numbers of bosons $N=20,28,32$
indicates the scaling behavior of \Eq{scaling1}).
The scaling exponent $\alpha$ depends on the initial state though, with
$\alpha=1/2$ for $|N,0\rangle$ and $\alpha=0.84$ for the initial state $|N/2,N/2\rangle$. 
\rmrk{Axes units are dimensionless.}
}
\label{fig:loc1}
\end{center}
\end{figure}

\subsection{Exact calculation - $\mathcal{N}_{\infty}$}

In order to include effects on large time scales, we calculated from $\mathcal{P}(t)$   
the participation number $\mathcal{N}_{\infty}$. It is directly calculated from the resolvent
via the expressions in \Eq{density00} and \Eq{r_density}. Some results for the initial state
$|N,0\rangle$ are depicted in \Fig{fig:loc0}, with a clear signature of a
transition, indicated by a maximum of $\mathcal{N}_{\infty}$ at ${1.8 \lesssim u_c \lesssim 1.9}$, 
with some weak $N$ dependence. From the semiclassical considerations, the transition should 
take place near the critical mean-field value ${u_c=2}$. 
On the other hand, looking at \Fig{fig:loc0} we see that 
\beq
\mathcal{N}_{\infty}(N,u)\sim 1 \ \ \ {\rm for}\ \ \ u\rightarrow\infty
\ .
\label{large_u}
\eeq
This is because in the soc-called Fock regime (${u\gg N^2}$) 
the preparations that we consider become eigenstates of the hamiltonian,
hence the participation number becomes unity.

In \Fig{fig:pn_b} we plot the normalized participation number $\mathcal{N}_{\infty}/\sqrt{N+1}$, 
as a function of the dimensionless parameter~$u$, for $|\psi_0\rangle=|N,0\rangle$ and different 
values of $N$. The curves fall on top of each other for small values
of $u$, whereas they depend on $N$ for larger values of $u$. 
In \Fig{fig:loc1} we focus \rmnt{removing "vicinity of $u=2$" which is somewhat misleading} 
on the range of $u$ values where the maximum appears, 
considering both preparations $|\psi_0\rangle=|N,0\rangle$ 
and $|\psi_0\rangle=|N/2,N/2\rangle$. 
We observe a scaling behavior with the scaling law
\beq
\mathcal{N}_{\infty}(N,u) \ \ \approx \ \ (N+1)^{\alpha} \ f(u) 
\ ,
\label{scaling1}
\eeq
\rmnt{fixed minor phrasing issues}
where the exponent $\alpha$ depends on the initial state. We have found ${\alpha\approx1/2}$ for $|N,0\rangle$
and ${\alpha\approx0.84}$ for $|N/2,N/2\rangle$. Also the scaling-function $f(u)$ depends on the
initial state: there is a characteristic maximum near ${u=1.8}$ for ${|\psi_0\rangle=|N,0\rangle}$ 
and near ${u=1.1}$ for ${|\psi_0\rangle=|N/2,N/2\rangle}$. 
Thus the exact results are qualitatively in accordance with the semiclassical expectation, 
but with some pronounced deviations, notably of $\alpha$ in the case of the Equator preparation.

\section{Discussion and Conclusion}

\rmnt{Below are the main textual changes. we have enough good stuff in this Ms, 
and there is no reason to take risks in using controversial semantics. Namely, 
most people will not regard self-trapping as quantum localization for obvious reasons.
See also my comment after Eq(39). So the of my suggested modification below is 
to avoid getting into semantic traps.}

The possibility of HSL in closed quantum systems has been discussed in general 
terms. We have extended the original ideas of Boltzmann and Heller, introducing 
a distinction between two notions of phase-space exploration. This provides 
naturally a semiclassical perspective for both weak and strong localization effects. 

In order to clarify the practical procedure for estimating the pertinent 
exploration measures, we have analyzed in detail the prototype Bosonic Josephson junction, 
where the number of bosons $N$ plays the role of inverse Planck constant. 
We have used both the semiclassical perspective and an exact quantum calculation
using the recursive projection method.  

We have explored the dependence of the participation number $\mathcal{N}_{\infty}$ 
on the number of bosons in the Josephson junction. Its use is convenient due to 
its direct connection with the return probability to the initial quantum state. 
We have found that its dependence on the dimensionless parameter~$u$ 
obeys the scaling laws \Eq{large_u} for $u\sim\infty$, 
and \Eq{scaling1} for ${0\lesssim u \lesssim 3}$, where the self-trapping transition takes place. 

Generally, the return probability as well as the participation number  
depend on the initial state $|\psi\rangle$. 
In the scaling law \Eq{scaling1}, the exponent changes from ${\alpha\approx 1/2}$ for 
${|\psi\rangle=|N,0\rangle}$ to ${\alpha\approx 0.84}$ for ${|\psi\rangle=|N/2,N/2\rangle}$.
The former value agrees with the naive semiclassical expectation, while the latter 
deviates significantly.
The quantum dynamics is more complex than the mean-field dynamics, as indicated by the examples
in \Fig{fig:jacobi} and \Fig{fig:dyn1}. This is related to the spectral properties of the 
participating eigenstates, namely, the appearance of fragmentation 
as the separatrix region is crossed (cf. \Fig{fig:dos}).
\rmnt{removed a few confusing sentences. In particular "in the quantum system, 
we cannot reproduce the $\log N$ behavior in the vicinity of the hyperbolic 
classically-unstable point, mentioned in Sect.\ref{sect:pendulum}". This is 
totally misleading because this prediction refers to coherent preparation that 
we did not consider in the present work. What we considered was Equator preparation, 
and the scaling was (roughly) as expected.}    

\rmnt{If we want regard to self-trapping as HSL, here is a suggestion 
for a careful non-controversial phrasing of this idea}
The self-trapping behavior of the mean-field approximation might be regarded  
as some kind of HSL. Considering a North Pole preparation (all the particles 
are initially in the left site), ignoring the possibility of tunneling, 
or breaking a bit the mirror symmetry of the Hamiltonian,
then for ${u>2}$ only half of Hilbert space is explored (${\mathcal{F}=1/2}$).  
Irrespective of that, there are other characteristic features of the quantum behavior,
such as the complex dynamics in \Fig{fig:dyn1}, the change of the spectral properties 
in \Fig{fig:dos}, and the shift of the transition point~$u_c$ in \Fig{fig:loc1}, which 
indicate a complex behavior, that cannot be anticipated by a simple mean-field approximation.

\ \\
{\bf Acknowledgements.-- }
This research has been supported by the Israel Science Foundation (grant No. 29/11).
One of the authors (V.I.Y) acknowledges financial support from the RFBR
(grant No. 14-02-00723)

\ \\

\appendix

\section{Spatial localization}
\label{app:P}

In this Appendix, we briefly recall how the occurrence of spatial 
localization can be quantified. 
A closed quantum system is defined by a Hamiltonian~$H$. Then its dynamics 
can be characterized by the transition amplitude $\langle n'|e^{-iHt}|n\rangle$, 
which describes the evolution of the initial state $|n\rangle$ over the 
time period $t$ and measures the overlap of the resulting state with the 
state $|n'\rangle$. The corresponding transition probability reads 
$P_t(n'|n)=|\langle n|e^{-iHt}|m\rangle|^2$, which is an observable 
quantity. An example for the latter is diffusion in a $d$--dimensional 
real space:
\beq
P_t(r'|r) \ \equiv \ |\langle r'|e^{-iHt}|r\rangle|^2=\frac{e^{-|r-r'|^2/4Dt}}{(4\pi Dt)^{d/2}}
\ ,
\label{tr_prob}
\eeq
which describes the probability to find a particle at site $r'$ after a 
period of time $t$, when it started from the site $r$. In particular, the 
probability for a particle to return to its starting point $r$ after time $t$ 
is given by the return probability
\beq
\mathcal{P}(t) \ = \ P_t(r|r) \ = \ \frac{1}{(4\pi Dt)^{d/2}}
\ ,
\eeq
which vanishes like the power law $t^{-d/2}$ for long times $t$. In other 
words, the particles diffuse further and further away from the starting point. 
Diffusion is a concept which has been realized in nature for classical particles 
and waves (e.g. for light). Quantum particles and other wave-like states can 
escape from diffusion if there is sufficient random scattering. In his seminal 
work, Anderson \cite{anderson58} suggested that quantum particles can localize 
in the presence of random scattering due to interference effects. It means that 
the particle stays in the vicinity of the initial site for all times. In more 
formal terms, the transition probability decays exponentially in space and is 
characterized by the localization length $\xi$,
\beq
P_t(r'|r) \ \sim \ P_{\infty} \, e^{-|r-r'|/\xi}
\eeq
for $t\sim\infty$, where the return probability $\mathcal{P}(t)$ is just the 
constant $P_{\infty}$. Thus, diffusion and localization can be characterized by the 
return probability. This quantity either vanishes on long time scales for 
diffusion or is nonzero for localization. It is convenient to integrate over 
all times to obtain the inverse participation number from the expression
\beq
\mathcal{N}_{\infty}^{-1} \ = \ \lim_{\epsilon\to0}\epsilon\int_0^\infty  
\mathcal{P}(t) \, e^{-\epsilon t}dt
\ \equiv \ P_{\infty} \; , 
\label{loc_p}
\eeq
which is either zero (diffusion) or nonzero (localization).

\section{Localization and entropy}
\label{app:S}

A localized quantum state is expected to display a smaller entropy than an 
extended delocalized state. Therefore the degree of delocalization can be 
characterized by the Shannon entropy or information entropy
\beq
S_I = - \sum_n p_n \ln p_n \;   ,
\eeq
where $p_n \equiv \langle n |\hat{\rho}| n \rangle$ is a diagonal matrix 
element of the statistical operator \cite{Wehrl_91,Mirbach_98}. The matrix 
elements can be taken with respect to any basis, for instance, the basis 
formed by the eigenvectors of the system Hamiltonian. Since the Shannon 
entropy can be treated as obtained from the von Neumann entropy by cancelling 
the off-diagonal matrix elements, it can be termed the diagonal entropy. For 
the basis of coherent states, the above form of the Shannon entropy was 
introduced by Wehrl \cite{Wehrl_78} and studied in a number of papers, 
e.g., \cite{Zyczkowski_90,Anderson_93,Buzek_95,Buzek_1995} and using other 
natural bases in Refs. \cite{Gorin_97,Frischat_97}. It is sometimes called 
the Wehrl entropy. The information entropy, with the time averaged statistical 
operators, has also been used as a measure of localization \cite{Thiele_84}. 
The diagonal entropy $S_I$ was shown to be useful for considering 
thermalization and equilibration of finite quantum systems \cite{Polkovnikov_11}. 
Note that the summation over $n$ in the information entropy can be replaced 
by the appropriate integration, when necessary. 

A generalized form of the information entropy is the Renyi entropy
\beq
 S_\gamma = \frac{1}{1-\gamma} \; \ln \;\sum_n p_n^\gamma \qquad
( 0 < \gamma < \infty ) \;  ,
\eeq
which has also been employed for quantifying localization \cite{Mirbach_98}. 
Localization was shown to be connected to entanglement entropy \cite{Serbin_15}.

\begin{figure}[t!]
\begin{center}
\includegraphics[width=6cm]{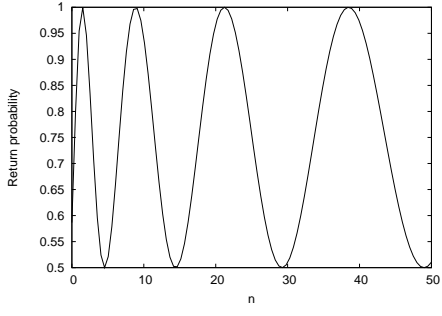}
\caption{
The partition number $\mathcal{N}_{\infty}$ of the Jaynes-Cummings model in \Eq{r_jc} for $g=\delta$.
}
\label{fig:jcm}
\end{center}
\end{figure}

\section{Jaynes-Cummings model}
\label{app:jcm}

The Jaynes-Cummings model describes cavity photons which interact with a 
two-level atom, where the latter can absorb or emit a single photon. It is 
defined by the Hamiltonian
\beq
H_{JC}=\hbar\omega(a^\dagger a+\frac{1}{2}\sigma_z)+\frac{\hbar\delta}{2}\sigma_z
+\frac{\hbar g}{2}(a\sigma_+ +a^\dagger \sigma_-) \; ,
\label{jcm_ham}
\eeq
which is acting on the product states 
$|n,\sigma\rangle=|n\rangle\otimes|\sigma\rangle$,
where $|n\rangle$ is a state of $n$ cavity photons and $|\sigma\rangle$ is a 
state of a two level system (e.g., an atom) with the atomic ground state 
$|\dn\rangle$ and the atomic excitation state $|\up\rangle$. The Pauli matrices 
$\sigma_{x,y,z}$ are operating on the two atomic levels, where 
$\sigma_\pm =\sigma_x\pm i\sigma_y$ creates and annihilates an excitation, 
respectively. The terms with $\sigma_z$ represent the energy splitting between 
the ground state and the excited state. The photon creation (annihilation) 
operators $a^\dagger$ ($a$) act only on the photon states. Finally, the 
coupling strength between the two-level atom and the cavity photons is $g$.

The Hamiltonian \Eq{jcm_ham} maps the state $|n,\up\rangle$ to a linear 
combination of $|n,\up\rangle$ and $|n+1,\dn\rangle$, and the state $|n,\dn\rangle$, 
to a linear combination of $|n,\dn\rangle$ and $|n-1,\up\rangle$. This implies 
that the eigenstates $|E_\pm,n\rangle$ are the linear combinations of two 
states with
\beq
\langle E_+,n|n,\up\rangle &=& \cos\alpha_n \ , \\
\langle E_-,n|n,\up\rangle &=& -\sin\alpha_n \ , \\
\langle E_+,n|n+1,\dn\rangle &=& \sin\alpha_{n} \ , \\ 
\langle E_-,n|n+1,\dn\rangle &=& \cos\alpha_{n} \; \ ,
\eeq
where
\beq
\alpha_n=\arctan \left(g\sqrt{n+1}/\delta\right) \; .
\eeq
This gives for the participation number (inverse return probability), with 
the initial states $|n,\up\rangle$ and $|n,\dn\rangle$,
\beq
\mathcal{N}_{\infty;n,\up}=\mathcal{N}_{\infty;n+1,\dn}=\cos^4\alpha_n+\sin^4\alpha_n 
\ ,
\label{r_jc}
\eeq
indicating HSL. This is plotted as a function of $n$ in \Fig{fig:jcm}.
This rather simple example demonstrates that the return probability 
$\mathcal{N}_{\infty}$ does not need to go to a simple asymptotic value for 
a large number of particles but can have an oscillatory behavior.

\section{Spin representation of the 2-site Bose-Hubbard model}
\label{app:spin}

The Hamiltonian $\mathcal{H}_{\text{BH}}$ can also be expressed as a 
nonlinear $SU(2)$ spin Hamiltonian in $N+1$ dimensions, when we write for 
the $SU(2)$ spin components
\beq
L_x &=& {1\over2}(a_1^\dagger a_2 + a_2^\dagger a_1) ,\\
L_y &=& {-i\over2}(a_1^\dagger a_2 - a_2^\dagger a_1) ,\\
L_z &=& {1\over2}(a_1^\dagger a_1-a_2^\dagger a_2) \ .
\label{spincomponents}
\eeq
Thus, the Hamiltonian in \Eq{ham00} becomes the nonlinear spin 
Hamiltonian in \Eq{spin_ham}.

\section{Double well without interaction}
\label{app:double}

For the expression in \Eq{rprob0}, the specific case of \Eq{ret_prob2} gives
\beq \nonumber
\mathcal{N}_{\infty}^{-1} 
&=& \lim_{\epsilon\to 0}\epsilon\int_0^\infty \cos^{2N}(Jt/2)e^{-\epsilon t}dt
\\ \nonumber
&=& 2^{-2N}\sum_{k=0}^{2N}{2N \choose k}\lim_{\epsilon\to 0}\epsilon\int_0^\infty e^{-[\epsilon+iJ(N-k)]t}dt
\\
&=& 2^{-2N}{2N \choose N}
\ \ \approx \ \ \sqrt{\frac{2}{\pi}}\frac{1}{\sqrt{N}}
\ ,
\eeq
where the approximation is the Stirling formula.

\clearpage


\begin{thebibliography}{99}

\bibitem{Blaizot_86}
J.P. Blaizot and G. Ripka,
{\it Quantum Theory of Finite Systems} 
(Massachusetts Insitutute of Technolodgy, Cambridge, 1986). 

\bibitem{cqed}
S. Haroche, J.M. Raimond, 
{\it Exploring the Quantum: Atoms, Cavities and Photons} 
(Oxford University Press, Oxford, 2006).

\bibitem{Walther_06}
H. Walther, B.T.H. Varcoe, B.G. Englert, and T. Becker,
Rep. Prog. Phys. {\bf 69}, 1325 (2006).

\bibitem{Lipparini_08}
E. Lipparini,
{\it Modern Many-Particle Physics} 
(World Scientific, Singapore, 2008).

\bibitem{Pethick_08}
C.J. Pethick and H. Smith,
{\it Bose-Einstein Condensation in Dilute Gases} 
(Cambridge University, Cambridge, 2008).

\bibitem{Yukalov09}
V.I. Yukalov, 
Laser Phys. {\bf 19}, 1 (2009).

\bibitem{Katsnelson_12}
M.I. Katsnelson,
{\it Graphene: Carbon in Two Dimensions}
(Cambridge University, Cambridge, 2012).

\bibitem{Birman_13}
J.L. Birman, R.G. Nazmitdinov, and V.I. Yukalov,
Phys. Rep. {\bf 526}, 1 (2013).

\bibitem{PSSV_11}
A. Polkovnikov, K. Sengupta, A. Silva, and M. Vengalatore, 
Rev. Mod. Phys. {\bf 83}, 863 (2011). 

\bibitem{Yukalov_11}
V.I. Yukalov,
Laser Phys. Lett. {\bf 8}, 485 (2011).

\bibitem{Williams_98}
C.P. Williams and S.H. Clearwater,
{\it Explorations in Quantum Computing} (Springer, New York, 1998).

\bibitem{Nielsen_00}
M.A. Nielsen and I.L. Chuang,
{\it Quantum Computation and Quantum Information} 
(Cambridge University, New York, 2000). 

\bibitem{Vedral_02}
V. Vedral,
Rev. Mod. Phys. {\bf 74}, 197 (2002).

\bibitem{Keyl_02}
M. Keyl,
Phys. Rep. {\bf 369}, 431 (2002).

\bibitem{anderson58}
P.W. Anderson, 
Phys. Rev. {\bf 109}, 1492 (1958).

\bibitem{Basko_06}
D.M. Basco, I.L. Aleiner, and B.L. Altshuler,
Ann. Phys. (N.Y.) {\bf 321}, 1126 (2006).

\bibitem{gora}
I.L. Aleiner, B.L. Altshuler, G.V. Shlyapnikov,
Nature Phys. {\bf 6}, 900 (2010). 

\bibitem{Pal_10}
A. Pal and D. Huse,
Phys. Rev. B {\bf 82}, 174411 (2010).

\bibitem{Huse_13}
D.A. Huse, R. Nandkishore, V. Oganesyan, A. Pal, and S.L. Sondhi,
Phys. Rev. B {\bf 88}, 014206 (2013).


\bibitem{QKRc}
\rmrk{G. Casati, B.V. Chirikov, F.M. Izrailev, and J. Ford,} 
in Stochastic Behaviour in classical and Quantum Hamiltonian Systems, 
Vol. 93 of Lecture Notes in Physics, 
edited by G. Casati and J. Ford (Springer, N.Y. 1979), p. 334.

\bibitem{QKRf}
\rmrk{S. Fishman, D.R. Grempel, and R.E. Prange,} 
Phys. Rev. Lett. 49, 509 (1982). 

\bibitem{TIPb}
\rmrk{F. Borgonovi and D.L. Shepelyansky,}
Nonlinearity 8, 877 (1995).

\bibitem{Kramer}
\rmrk{B. Kramer and A. MacKinnon},
Rep. Prog. Phys. {\bf 56}, 1469 (1993). 

\bibitem{Modugno}
\rmrk{G. Modugno},
Rep. Prog. Phys. {\bf 73}, 102401 (2010). 




\bibitem{TIPs}
\rmrk{D. L. Shepelyansky,}
Phys. Rev. Lett. 73, 2607 (1994).


\bibitem{TIPi}
\rmrk{Y. Imry,}
Europhysics Letters 30, 405 (1995).

\bibitem{ckt}
C. Khripkov, D. Cohen, and A. Vardi, 
Phys. Rev. E {\bf 87}, 012910 (2013). 

\bibitem{Heller_87}
E.J. Heller,
Phys. Rev. A {\bf 35}, 1360 (1987).

\bibitem{dittrich}
T. Dittrich,
Phys. Rep. {\bf 271}, 267 (1996).
 
\bibitem{brk}
D. Cohen,
J. Phys. A {\bf 31}, 277 (1998).

\bibitem{lea1} 
E.J. Torres-Herrera, L.F. Santos,
Phys. Rev.B {\bf 92}, 014208 (2015).

\bibitem{lea2}
L. F. Santos and F. Perez-Bernal,
arXiv:1506.06765 (2015).

\bibitem{sfc}
G. Arwas, A. Vardi, and D. Cohen,
Sci. Rep. {\bf 5}, 13433 (2015).   

\bibitem{scar1}
L. Kaplan and E.J. Heller, 
Phys. Rev. E {\bf 59}, 6609 (1999).

\bibitem{scar2}
L. Kaplan, 
Nonlinearity {\bf 12}, R1 (1999).

\bibitem{Logan_90}
D.E. Logan and P.G. Wolynes,
J. Chem. Phys. {\bf 93}, 4994 (1990). 

\bibitem{ziegler03}
K. Ziegler, 
Phys. Rev. A {\bf 68}, 053602 (2003).

\bibitem{ziegler11}
K. Ziegler, 
J. Phys. B: At. Mol. Opt. Phys. {\bf 44}, 145302 (2011).

\bibitem{jaynes63}
E.T. Jaynes and F.W. Cummings, 
Proc. Inst. Elect. Eng. {\bf 51}, 89 (1963).

\bibitem{cummings65}
F.W. Cummings, 
Phys. Rev. {\bf 140}, A1051 (1965).

\bibitem{milburn97}
G.J. Milburn, J. Corney, E.M. Wright, D.F. Walls, 
Phys. Rev. A {\bf 55}, 4318 (1997).

\bibitem{ketterle04}
Y. Shin, M. Saba, A. Schirotzek, T.A. Pasquini, A.E. Leanhardt, D.E. Pritchard, 
and W. Ketterle,
Phys. Rev. Lett. {\bf 92}, 150401 (2004).

\bibitem{oberthaler05}
M. Albiez, R. Gati, J. F\"olling, S. Hunsmann, M. Cristiani, and M.K. Oberthaler, 
Phys. Rev. Lett. {\bf 95}, 010402 (2005).

\bibitem{Boukobza}
E. Boukobza, M. Chuchem, D. Cohen and A. Vardi,
Phys. Rev. Lett. {\bf 102}, 180403 (2009). 

\bibitem{gati06}
R. Gati et al., 
New J. Phys. {\bf 8}, 189 (2006).

\bibitem{oberthaler07}
R. Gati and M.K. Oberthaler, 
J. Phys. B: At. Mol. Opt. Phys. {\bf 40} (2007) R61.

\bibitem{cohen10}
M. Chuchem, K. Smith-Mannschott, M. Hiller, T. Kottos, A. Vardi and D. Cohen,
Phys. Rev. A {\bf 82}, 053617 (2010).

\bibitem{eilbeck85}
J.C. Eilbeck, P.S. Lomdahl, and A.C. Scott, 
Physica D {\bf 16}, 318 (1985).

\bibitem{abramowitz}
M. Abramowitz and I. Stegun, 
{\it Handbook of Mathematical Functions} (Dover, New York, 1972).

\bibitem{YYB_97}
V.I. Yukalov, E.P. Yukalova, and V.S. Bagnato,
Phys. Rev. A {\bf 56}, 4845 (1997).

\bibitem{YYB_00}
V.I. Yukalov, E.P. Yukalova, and V.S. Bagnato,
Laser Phys. {\bf 10}, 26 (2000).

\bibitem{YYB_02}
V.I. Yukalov, E.P. Yukalova, and V.S. Bagnato,
Phys. Rev. A {\bf 66}, 043602 (2002).


\bibitem{Wehrl_91}
A. Wehrl,
Rep. Math. Phys. {\bf 30}, 119 (1991). 

\bibitem{Mirbach_98}
B. Mirbach and H.J. Korsch,
Ann. Phys. (N.Y.) {\bf 265}, 80 (1998).

\bibitem{Wehrl_78}
A. Wehrl,
Rev. Mod. Phys. {\bf 50}, 221 (1978).

\bibitem{Zyczkowski_90}
K. Zyczkowski,
J. Phys. A {\bf 23}, 4427 (1990).

\bibitem{Anderson_93}
A. Anderson and J.J. Halliwel,
Phys. Rev. D {\bf 48}, 2753 (1993).

\bibitem{Buzek_95}
V. Bu\v{z}ek, C.H. Keitel, and P.L. Knight,
Phys. Rev. A {\bf 51}, 2575 (1995).

\bibitem{Buzek_1995}
V. Bu\v zek, C.H. Keitel, and P.L. Knight,
Phys. Rev. A {\bf 51}, 2594 (1995).

\bibitem{Gorin_97}
T. Gorin, H.J. Korsch, and B. Mirbach,
Chem. Phys. {\bf 217}, 147 (1997).

\bibitem{Frischat_97}
S.D. Frischat and E. Doron,
J. Phys. A {\bf 30}, 3613 (1997).  

\bibitem{Thiele_84}
E. Thiele and J. Stone,
J. Chem. Phys. {\bf 80}, 5187 (1984).

\bibitem{Polkovnikov_11}
A. Polkovnikov,
Ann. Phys. (N.Y.) {\bf 326}, 486 (2011).

\bibitem{Serbin_15}
M. Serbin, Z. Papi\'{c}, and D.A. Abanin, 
arXiv:1507.01635 (2015).

\bibitem{ArQ1}
D. M. Leitner and P. G. Wolynes,
Phys. Rev. Lett. {\bf 79}, 55 (1997).

\bibitem{ArQ2}
V.Ya. Demikhovskii, F.M. Izrailev, and A.I. Malyshev, 
Phys. Rev. Lett. {\bf 88}, 154101 (2002).


\end{thebibliography}
\end{document}